\documentstyle[epsfig]{aipproc}

\newcommand{\bec}[1] {\begin{equation}\label{#1} }
\newcommand{\eec} {\end{equation} }
\begin{document}

\begin{flushright}
UR-1577\\
ER/40685/936\\
hep-ph/9907359\\
July 1999 \\
\end{flushright}

\title{Lepton - Chargino Mixing and R-Parity Violating SUSY
\thanks{Presented by C.~Macesanu at 21th Annual MRST (Montreal-Rochester-Syracuse-Toronto) Meeting 
  on High-Energy Physics, Ottawa, 10-12 May 1999. }
}

\author{Mike Bisset}
\address{Department of Physics, Tsinghua University, Beijing, 100084, China}
\author{Otto C.W. Kong, Cosmin Macesanu and Lynne H. Orr}
\address{Department of Physics and Astronomy,
University of Rochester,Rochester, New York 14627}

\maketitle

\begin{abstract}
We present a study of charged lepton mass matrix diagonalization
in R-parity violating SUSY. The case in which the  bilinear 
couplings $\mu_i$ have large values is given
special attention.
\end{abstract}

\section*{Introduction}
R-parity violating SUSY is a subject which has enjoyed a lot of interest in the
past few years. The possibility of mixing between particles and superpartners,
allowed in these models, makes for very interesting phenomenology.
However, due to the large number of parameters present, this subject
 is also quite
difficult to study. Recently, it has been found 
(see \cite{BKMO1}) that by working in a specific
basis (single VEV parametrization) the analysis of the fermionic sector is
greatly simplified, without loss of generality.
 In this basis, the only non-MSSM parameters that 
play a role in the leptonic phenomenology at tree level are the three
RPV bilinear couplings $\mu_i$.

The phenomenological consequences
% of the mixing
%between leptons and charginos  
of this model in the fermionic sector have been 
extensively studied \cite{BKMO2}.
 This paper aims to detail the technical aspects concerning the 
chargino-lepton mass matrix diagonalization. It has been found that 
for large values of the couplings $\mu_i$ (of order of hundreds GeV and
above) this problem is not trivial. We should point out that 
our interest in this issue {\it here} is mostly
theoretical: most of the range of $\mu_i$ values relevant to the discussion 
below is not allowed by 
the experimental constraints (except in special cases; 
for more details on this subject see \cite{BKMO2}).

 The framework is the same
as in \cite{BKMO1}.  The mass matrix can be written:

$$
M_c = \left[
\begin{array}{ccccc}
M_2 & g_v & 0 & 0 & 0 \\
g_v' & \mu_0 &		0 & 	 0    & 0 \\
 0   & \mu_1 & \bar{m}_1 &  0  & 0    \\
 0   & \mu_2 &	0 & \bar{m}_2 & 0   \\
 0   & \mu_3 & 	0 & 0 & \bar{m}_3 
\end{array}
\right]
$$
where $g_v=g_2 v_u/\sqrt{2} =\sqrt{2} M_w \mbox{sin}\beta,\ 
g_v'=g_2 v_d/\sqrt{2} = \sqrt{2} M_w \mbox{cos}\beta$. Here, by $\bar{m}_i$
we denote the Yukawa masses of the three leptons $e,\mu,\tau$. The physical
masses will be denoted by $m_i$.

The aim is to go from the weak interaction fields, in terms of which the 
lagrangian is written initially, to the mass eigenstate (physical) fields,
which can be observed experimentally. To this purpose, we rotate the left
fields by a matrix $U^L$ and the right fields by a matrix $U^R$. These 
rotations 
will also diagonalize the squared mass matrices:
$$ U^{L\dagger}\ M^L \ U^L = U^{R\dagger}\ M^R \ U^R =\
\mbox{diag} \{ M_{\chi_1}^2,M_{\chi_2}^2,m_i^2 \}
$$
where $M^L=M_c^{\dagger} M_c,\ M^R=M_c M_c^{\dagger}$.

The mixing between leptons and charginos naturally leads to changes in
the couplings of these particles to the gauge bosons (see \cite{NowaP}).
In the case of $Z$ coupling we will have :
$$A_{ij}^L=\tilde{A}_{ij}^L + (1-2\mbox{sin}^2\theta_W) \delta_{ij}, \ \
\tilde{A}_{ij}^L = U^L_{i1}\ U^L_{j1}
$$
$$A_{ij}^R=\tilde{A}_{ij}^R -2 \mbox{sin}^2\theta_W \delta_{ij}, \ \
\tilde{A}_{ij}^R = 2 U^R_{i1}\ U^R_{j1} + U^R_{i2}\ U^R_{j2}
$$
The diagonal ($\delta_{ij}$) terms are the SM values, while the 
$\tilde{A}$ terms are consequences of the mixing. Being nondiagonal, 
they lead to anomalous $Z$ couplings and nonstandard decays 
({\it e.g.} $Z \rightarrow e\mu, \ \mu \rightarrow eee$). It turns out that,
in most cases, 
the anomalous left coupling is the important one; the anomalous right
coupling is proportional with the product $m_i m_j$ and very small numerically.
In what follows we will concentrate on the left rotation matrix. 

\section{Analytic diagonalization of mass matrix}

In this section we will analyze approximate analytical solutions to
our diagonalization problem. The perturbative solution for small $\mu$'s has been given in \cite{BKMO1}; we present it here for comparision with further results:
\bec{seesaw}
 U^L_{i1}=\mu_i \frac{\sqrt{2} M_W \hbox{cos}\beta}{\Delta} \ \ , 
\ \ i={e,\mu,\tau}
\eec 
$\Delta=\mu_0 M_2 - g_v g_v'$.
%Although the seesaw approximation used here has a limited range of validity,
This formula reveals to us one important fact:  the strength 
of the mixing decreases at large tan$\beta$. The phenomenological consequences
of this behaviour have been analyzed in \cite{BKMO1},\cite{BKMO2}.

 The fact that the above formula doesn't work at
large $\mu$ is apparent; indeed, as we increase $\mu_i$, the components
 $U^L_{i1}$, as given by the above formulae, increase indefinitely, which is not allowed by $U$ matrix unitarity.

It is possible to diagonalize the matrix without assuming that the 
$\mu_i$ are small. Instead, we will take the Yukawa masses to be small; 
actually, we will take them to be zero in the matrix $M^L$. Then, we get the 
following solution: 
\bec{umat}
U^L=\left[
\begin{array}{ccccc}
x_1    & x_1'   & \mu_1 \frac{g_v'}{\Delta_e} 
                & \mu_2 \frac{g_v' \Delta}{\Delta_{\mu} \Delta_e}
                & \mu_2 \frac{g_v' \Delta}{\Delta_{\tau} \Delta_{\mu}} \\
x_2    & x_2'   & -\mu_1 \frac{M_2}{\Delta_e} 
                & -\mu_2 \frac{M_2 \Delta}{\Delta_{\mu} \Delta_e}
                & -\mu_2 \frac{M_2 \Delta}{\Delta_{\tau} \Delta_{\mu}} \\
x \frac{\mu_1}{\mu_5} & x' \frac{\mu_1}{\mu_5}  
	& \frac{\Delta}{\Delta_e}
	& -\frac{\mu_2}{\Delta_{\mu}} \frac{\mu_1 \alpha_2}{\Delta_e}
	& -\frac{\mu_3}{\Delta_{\tau}} \frac{\mu_1 \alpha_2}{\Delta_{\mu}} \\
x \frac{\mu_2}{\mu_5} & x' \frac{\mu_2}{\mu_5}
	& 0
	& \frac{\Delta_e}{\Delta_{\mu}}
	& -\frac{\mu_3}{\Delta_{\tau}} \frac{\mu_2 \alpha_2}{\Delta_{\mu}} \\
x \frac{\mu_3}{\mu_5} & x' \frac{\mu_3}{\mu_5}
	& 0
	& 0
	& \frac{\Delta_{\mu}}{\Delta_{\tau}}
\end{array}
\right]
\eec
where $\Delta_e^2=\Delta^2 + \alpha_2 \mu_1^2, 
\Delta_{\mu}^2=\Delta^2 + \alpha_2 (\mu_1^2 + \mu_2^2)$ , 
$\Delta_{\tau}^2=\Delta^2 + \alpha_2 (\mu_1^2 + \mu_2^2 + \mu_3^2)$, and
$\alpha_2=M_2^2 + g_v'^2$. The first two columns correspond to the two charginos, while the last three columns correspond to the leptons ($ e, \mu
, \tau$ respectively).

Note that, in the approximation used, the physical masses of the particles
(eigenvalues of the matrix $M^L$) are also zero; as a consequence, the three
lepton eigenvectors are degenerate. To get the correct combination, we employ
a limiting procedure: start from the exact eigenvectors, and let the 
electron mass, muon mass, and tau mass go to zero, in this order. 
If the lepton mass hierarchy holds also for the Yukawa masses
($\bar{m}_e << \bar{m}_{\mu} << \bar{m}_{\tau}$) it can
be shown that this way we get the correct result (\ref{umat}).
 
 \vspace{0.3cm}
 Besides lepton eigenvectors, other quantities of interest obtained
through diagonalization of the mass matrix are chargino masses. In the 
small Yukawa mass approximation, we get
$$ M_{\chi_{1,2}} = \frac{\alpha_1 + \alpha_2}{2} \pm 
  \sqrt{(\alpha_1 - \alpha_2)^2 + 4(M_2 g_v + \mu_0 g_v')^2} $$
with $\alpha_2= \mu_0^2 + g_v^2 + \mu_5^2$
($\mu_5^2=\mu_1^2+\mu_2^2+\mu_3^2$). Interpretation of the quantities
$\alpha_1 $ and $\alpha_2$ is straightforward; at large $\mu_5$, the mass
of the heavier chargino is $M_{\chi_1}\approx \sqrt{\alpha_1} \approx \mu_5$,
while the mass of the lighter chargino is $M_{\chi_2}\approx \sqrt{\alpha_2}$.
Actually, it can be shown that the lighter chargino mass increases 
monotonically from the MSSM value (for $\mu_5 = 0$) to $\sqrt{\alpha_2}$
(for $\mu_5 \rightarrow \infty$).
 This behaviour has important phenomenological consequences. 
 Consider the fact that the lower limit on 
the lighter chargino mass $M  {\chi_2} > 90 $ GeV eliminates part of the 
$(M_2,\mu_0)$ plane in the MSSM. With R-parity violating terms, you can
expect that this excluded 
region will shrink; indeed, if we make $\mu_5$ big enough,
it might potentially go away completely. The fact that $M  {\chi_2}$ is
limited above by $ \sqrt{\alpha_2}$ means that some region
in the $(M_2,\mu_0)$ plane does in fact remain excluded, no matter
how strong the R-parity is violated. 
This region is given by the equation:
$$  \sqrt{\alpha_2} < 90\ \hbox{GeV}$$
or, at large tan$\beta$, $M_2<90$ GeV. This result is supported by the 
exact numerical analysis presented in \cite{BKMO1}, \cite{BKMO2}.

\begin{figure}[t!] % fig 1
\centerline{\epsfig{file=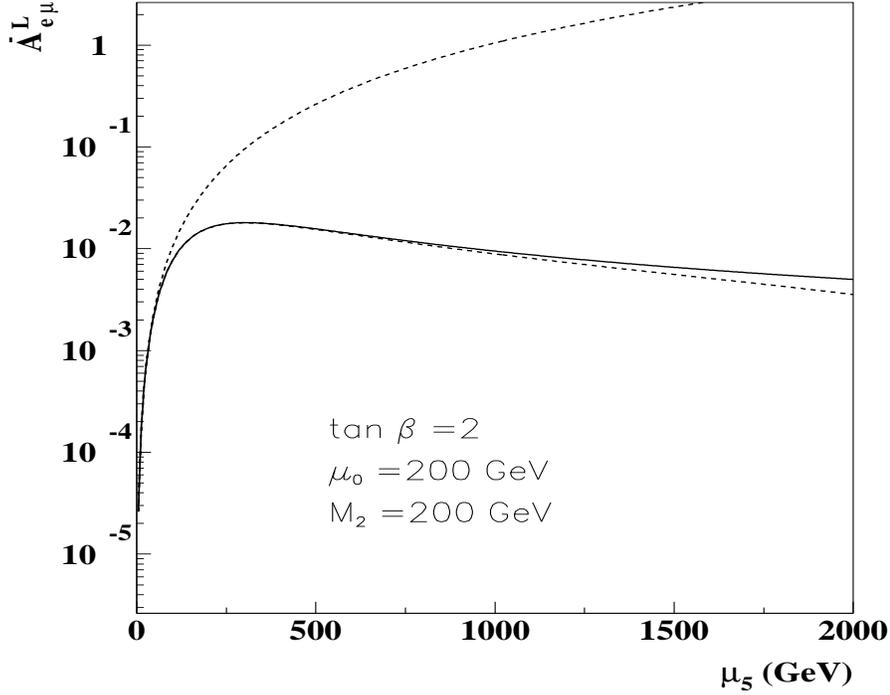,height=4.in,width=5.in}}
\vspace{10pt}
\caption{  Strength of left-handed $Ze\mu$ anomalous coupling : dotted line:
 small $\mu$ approximation (\ref{seesaw});
solid line: small Yukawa mass approximation (\ref{umat}); 
dashed line: exact numerical
result; $\mu_i$ ratio 1:1:1}
\label{fig1}
\end{figure}

\vspace{0.3cm}
Let's turn back to the lepton eigenvectors and consider the anomalous 
lepton-lepton-Z couplings. For simplicity, let's take the left-handed
Z$e\mu$ coupling:
\bec{lcoup}
 \tilde{A}^L_{e\mu}=\mu_1 \mu_2 \frac{g_v'^2 \ \Delta}
{\Delta_e^2 \ \Delta_{\mu}}
\eec 
At small $\mu$'s, the strength of the coupling increases with $\mu_1 \mu_2$, 
while at large $\mu$'s it decreases like $1/\sqrt{\mu_5^2}$ 
(see Fig. \ref{fig1}).
 In between it will
reach a maximum value:
$$ \tilde{A}^L_{e\mu \ max}= C \ \frac{g_v'^2}{M_2^2 + g_v'^2}$$
C being a constant which depends only on the ratio $\mu_1/\mu_2$. 
Experimental constraints on anomalous branching ratios or lepton number
violating decays (which can generally be written in the form
 $ \tilde{A}^L_{e\mu}<A_0$)  will then be satisfied not only in the region
 of  small $\mu_5$, but also for large values of $\mu_5$. Moreover, 
if in some region of parameter space 
 $ \tilde{A}^L_{e\mu \ max}  <A_0$  for some process, 
  then that particular process won't 
 contribute at all to constraints on $\mu_5$ values.

\vspace{0.3cm}
The analytic solution (\ref{umat}) 
derived in this section is not valid at arbitrarily large $\mu$'s.
The approach used to derive the lepton eigenvectors works only as long
as  $\bar{m}_e << \bar{m}_{\mu} << \bar{m}_{\tau}$. Once we get close
to the boundaries of the region where the diagonalization 
problem has solutions
(see next section) this relation does not hold anymore. However, the 
numerical results  for the anomalous couplings $\tilde{A}^L_{ij}$
show an even steeper decrease in this region than that given by  
(\ref{lcoup}). As a consequence, for sufficiently large $\mu_5$, 
constraints from electroweak processes like lepton or $Z$ decay 
dissapear. This region of large $\mu_5$ is excluded on the basis
of strong interaction processes
($\pi$ decay, or neutrinoless beta decay). However, the corresponding
case involving mainly large $\mu_3$ may be phenomenologically viable
\cite{BKMO2}.

\section{Yukawa masses}
To be able to perform an exact diagonalization of the mass matrix,
we first have to find the Yukawa masses of the leptons. Note that this
is not the standard problem, in which we have a matrix and we have to 
find eigenvalues. In this case, we know three of the eigenvalues (the
physical masses $m_i$ of the leptons) and we have to find some elements of the
matrix itself (the Yukawa masses $\bar{m}_i$).

 The direct approach would be to use the eigenvalue equations:
\bec{sis1}
 \mbox{det}(M - \lambda I ) =0 \ \ ,\ \ \lambda=m_e^2,m_{\mu}^2,m_{\tau}^2
\eec
 This is a three by three system of nonlinear equations, and is not easy
to solve even numerically (except in particular cases). So, we will use
another approach.

This approach is based on the observation that if we know the solution
for some values ${\mu_i^2}$, we can find the solution at an neighbouring
point ${\mu_i^2 + \mbox{d}\mu_i^2}$. For simplicity, let's consider the 
ratio of the $\mu$'s fixed, and their magnitude given by a parameter $t$:
$$\mu_i = r_i \sqrt{t}$$
 Write the eigenvalue equations:
$$E(t,\bar{m}_i^2,\lambda) = \mbox{det}(M - \lambda I ) =0 $$
To an infinitesimal modification in $t$ will correspond an infinitesimal 
modificaton in the Yukawa parameters $\bar{m}_i$:
\bec{sis2}
\frac{\partial E}{\partial t}\ +\ \frac{\partial E}{\partial  \bar{m}_i^2}\
\frac{\mbox{d} \bar{m}_i^2}{ \mbox{d} t}\ =\ 0
\eec
Now, we can numerically integrate this system of linear equations from zero
to whatever value of $t$ we need.

\vspace{0.3cm}
What about the existence of solutions for our problem? Let's suppose we
can solve the system (\ref{sis1});
 in order that the result make sense, we require
the solutions to be real. This will restrict the allowable values of
$\mu_i$ to some domain $D$ in $\mu_i$ space. What this means, in terms 
of solutions derived with the help of (\ref{sis2}), 
is that we can increase $t$ only
as long as we stay inside this domain. When we reach its boundary, usually
what happens is that the determinant of the system
 (\ref{sis2}) becomes zero, and
we cannot solve for $\mbox{d} \bar{m}_i$.

Another relevant question is if this domain is simply-connected; 
that is, starting
from $\mu_i = 0$, can we reach any point of it with a path formed by
connected  straight
lines? In other words, does integrating the system 
(\ref{sis2}) allow us access to all 
the solutions to (\ref{sis1})?

We do not know the answer to this question for the general case of three
leptons. But, if we consider the simpler case of only two leptons 
(presume one of the $\mu_i$ is zero), we can write (\ref{sis1})
in the form:
$$ 
\left\{
\begin{array}{c}   
a_1 \bar{m}_1^2 + a_2 \bar{m}_2^2 = s \\
a_1 a_2 \bar{m}_1^2 \bar{m}_2^2 =p 
\end{array}
\right.
$$
with solution:
$$ (a \bar{m}^2)_{1,2} = \frac{s}{2} \mp \sqrt{s^2 - 4p}$$ 
The quantity $A=s^2 - 4p$ becomes 0 for some value $\mu_5=\mu_{5\ max}$
(which gives us the boundary for the domain $D$),
and it can be shown that only for $\mu_5 < \mu_{5\ max}$ is $A$ positive
(necessary condition for real solutions). Note that $\mu_{5\ max}$ is
generally around a few TeV.

\vspace{0.3cm} 
Another interesting issue is the problem of lepton mass hierarchy in this model.
In the Standard Model (or MSSM) we have $\bar{m}_e << \bar{m}_{\mu}
<<\bar{m}_{\tau}$. These relations need not hold in our R-parity violating
scenario. Take, for example, the two lepton mixing case presented above.
If $\mu_1 = \mu_2$, then, at $ \mu_5=\mu_{5\ max}$, we have 
$\bar{m}_1=\bar{m}_2$ (this will happen at quite large $\mu_5$ values,
though; $\mu_{5\ max} \sim m_2/m_1$). If $\mu_1 > \mu_2$, it is even 
possible to get $\bar{m}_1$ greater than $\bar{m}_2$. The next question
is if this behaviour holds for the general case of three lepton mixing.
The possibility of finding points
in parameter space where the three Yukawa masses are of the same order
of magnitude (or maybe even equal) is particularly interesting. 
Unfortunately, the technical difficulties encountered in working with 
the nonlinear system (\ref{sis1}) have stopped us from getting an 
answer to this question so far. 

\section{Conclusions}
For large R-parity violating terms, the mixing between charginos and 
charged leptons has different characteristics than at small $\mu_i$. 
We have presented approximate analytical expressions for both regimes, 
which can help to understand numerical results. 
%The case of large $\mu_3$ is even phenomenologically viable.

The problem of existence of solutions of the system (\ref{sis1})
- finding the Yukawa masses so that three of the mass matrix 
eigenvalues will be equal to the physical lepton masses -
is still unsolved for the general case of three lepton mixing
(although within the phenomenologically viable region, numerical solutions
are always successfully obtained in \cite{BKMO2}).
For two lepton mixing it can be solved, and it 
has been shown that there are regions
in parameter space where the Yukawa masses of the two particles are
of the same order of magnitude.
%Concerning the solvability of the diagonalization problem: 
%The domain of 

\end{document}